\documentclass[]{an}
\usepackage{graphicx}
\usepackage{times}
\begin{document}
\clubpenalty10000
\sloppy
\Pagespan{000}{000}
\Yearpublication{2010}%
\Yearsubmission{2010}%
\Month{00}%
\Volume{0000}%
\Issue{00}%

\title{Asteroseismology of Solar-type Stars with Kepler I: Data
Analysis}
\author{C. Karoff\inst{1,2}\fnmsep\thanks{\email{karoff@bison.ph.bham.ac.uk}}, W.~J.~Chaplin\inst{1}, T.~Appourchaux\inst{3}, Y.~Elsworth\inst{1}, R.~A.~Garcia\inst{4}, G.~Houdek\inst{5}, T.~S.~Metcalfe\inst{6}, J.~Molenda-{\.Z}akowicz\inst{7}, M.~J.~P.~F.~G.~Monteiro\inst{8}, M.~J.~Thompson\inst{9}, J.~Christensen-Dalsgaard\inst{2}, R.~L.~Gilliland\inst{10}, H.~Kjeldsen\inst{2}, S.~Basu\inst{11}, T.~R.~Bedding\inst{12}, T.~L.~Campante\inst{2,8}, P.~Eggenberger\inst{13}, S.~T.~Fletcher\inst{14}, P.~Gaulme\inst{3}, R.~Handberg\inst{2}, S.~Hekker\inst{1}, M.~Martic\inst{15}, S.~Mathur\inst{6}, B.~Mosser\inst{16}, C.~Regulo\inst{17, 18}, I.~W.~Roxburgh\inst{19}, D.~Salabert\inst{17, 18}, D.~Stello\inst{12}, G.~A~Verner\inst{19}, K.~Belkacem\inst{20}, K.~Biazzo\inst{21}, M.~S.~Cunha\inst{8}, M.~Gruberbauer\inst{22}, J.~A.~Guzik\inst{23}, F.~Kupka\inst{24}, B.~Leroy\inst{16}, H.-G.~Ludwig\inst{25}, S.Mathis\inst{26}, A.~Noels\inst{20}, R.~W.~Noyes\inst{27}, T.~Roca Cortes\inst{17,18}, M.~Roth\inst{28}, K.~H.~Sato\inst{4}, J.~Schmitt\inst{29}, M.~D.~Suran\inst{30}, R.~Trampedach\inst{31}, K.~Uytterhoeven\inst{4}, R.~Ventura\inst{32} \& P.~A.~Wilson\inst{33,34}}
\titlerunning{Asteroseismology of Solar-Type Stars with Kepler I: Data Analysis}
\authorrunning{C. Karoff et al.}
\institute{
School of Physics and Astronomy, University of Birmingham, Edgbaston, Birmingham B15 2TT, UK 
\and 
Department of Physics and Astronomy, Aarhus University, Ny Munkegade 120, DK-8000 Aarhus C, Denmark 
\and 
Institut d'Astrophysique Spatiale, Universit\'{e} Paris XI-CNRS (UMR8617), Batiment 121, 91405 Orsay Cedex, France 
\and
Laboratoire AIM, CEA/DSM-CNRS, Universit\'e Paris 7 Diderot, IRFU/SAp-SEDI, Centre
de Saclay, 91191, Gif-sur-Yvette, France.
\and
Institute of Astronomy, University of Vienna, A-1180 Vienna, Austria
\and
High Altitude Observatory and Scientific Computing Division, National Center for Atmospheric Research, Boulder, CO 80307, USA
\and
Astronomical Institute, University of Wroclaw, ul. Kopernika, 11, 51-622 Wroclaw, Poland
\and
Centro de Astrof\'{\i}sica and DFA-Faculdade de Ci\^encias, Universidade do Porto, Ruas das Estrelas, 4150-762 Porto, Portugal
\and
School of Mathematics and Statistics, University of Sheffield, Hounsfield Road, Sheffield S3 7RH, UK
\and
Space Telescope Science Institute, Baltimore, MD 21218, USA
\and
Department of Astronomy, Yale University, P.O. Box 208101, New Haven, CT 06520-8101, USA
\and
Sydney Institute for Astronomy (SIfA), School of Physics, University of Sydney, NSW 2006, Australia
\and
Geneva Observatory, University of Geneva, Maillettes 51, 1290 Sauverny, Switzerland
\and
Materials Engineering Research Institute, Faculty of Arts, Computing, Engineering and Sciences, Sheffield Hallam University, Sheffield, S1 1WB, UK
\and
LATMOS-IPSL, CNRS, Universit\'{e} de Versailles Saint-Quentin, 11, boulevard d'Alembert, 78280 Guyancourt, France
\and
LESIA, CNRS, Universit\'{e} Pierre et Marie Curie, Universit\'{e}, Denis Diderot, Observatoire de Paris, 92195 Meudon Cedex, France 
\and
Departamento de Astrof\'{i}sica, Universidad de La Laguna, E-38206 La Laguna, Tenerife, Spain 
\and
Instituto de Astrof\'{i}sica de Canarias, E-38200 La Laguna, Tenerife, Spain
\and
Queen Mary University of London, Astronomy Unit, Mile End Road, London E1 4NS, UK
\and
D\'{e}partement d'Astrophysique, G\'{e}ophysique et Oc\'{e}anographie (AGO), Universit\'{e} de Li\'{e}ge, All\'{e}e du 6 Ao\"{u}t 17 4000 Li\'{e}ge 1, Belgium
\and
Arcetri Astrophysical Observatory, Largo Enrico Fermi 5, 50125 Firenze, Italy 
\and
Department of Astronomy and Physics, Saint Mary's University, Halifax, NS B3H 3C3, Canada
\and
Los Alamos National Laboratory, Los Alamos, NM 87545-2345, USA
\and
Faculty of Mathematics, University of Vienna, Nordbergstra{\ss}e 15, A-1090 Wien, Austria 
\and
ZAH -- Landersternwarte, K\"onigstuhl 12, 69117 Heidelberg, Germany
\and
Laboratoire AIM, CEA/DSM-CNRS-Universit\'e Paris Diderot, IRFU/SAp Centre de Saclay,
F-91191 Gif-sur-Yvette, France
\and
Smithsonian Astrophysical Observatory, Cambridge, MA
\and
Kiepenheuer-Institut f\"ur Sonnenphysik, Sch\"oneckstr. 6, 79104 Freiburg, Germany
\and
Observatoire de Haute-Provence, F-04870, St.Michel lÕObservatoire, France  
\and
Astronomical Institute of the Romanian Academy, Str. Cutitul de Argint, 5, RO 40557, Bucharest, Romania 
\and
JILA, University of Colorado, 440 UCB, Boulder, CO 80309-0440, USA 
\and
INAF Osservatorio Astrofisico di Catania, Via S.Sofia 78, 95123, Catania, Italy 
\and
Nordic Optical Telescope, Apartado 474, E-38700 Santa Cruz de la Palma, Santa Cruz de Tenerife, Spain
\and
Institute of Theoretical Astrophysics, University of Oslo, P.O. Box 1029, Blindern, N-0315 Oslo, Norway 
}
\received{??}
\accepted{??}
\publonline{later}
\keywords{stars: interiors, stars: late-type, stars: oscillations}
\abstract{We report on the first asteroseismic analysis of solar-type
stars observed by Kepler. Observations of three G-type stars, made at
one-minute cadence during the first 33.5d of science operations,
reveal high signal-to-noise solar-like oscillation spectra in all
three stars: About 20 modes of oscillation can clearly be
distinguished in each star. We discuss the appearance of the
oscillation spectra, including the presence of a possible signature of
faculae, and the presence of mixed modes in one of the three stars.}
\maketitle

\section{Introduction}
The year 2009 marked an important milestone in asteroseismology, with
the launch of the NASA \emph{Kepler} Mission (Gilliland et
al. 2010). \emph{Kepler} will realize significant advances in our
understanding of stars, thanks to its asteroseismology program,
particularly for cool (solar-type) main-sequence and subgiant stars
that show solar-like oscillations, i.e., small-amplitude oscillations
intrinsically damped and stochastically excited by the near-surface
convection (see Christensen-Dalsgaard 2004 for a recent review). Solar-like oscillation
spectra have many modes excited to observable amplitudes. The rich
information content of these seismic signatures means that the
fundamental stellar properties (e.g., mass, radius, and age) may be
measured and the internal structures constrained to levels that would
not otherwise be possible (e.g., see Gough 1987; Cunha et al. 2007).

For its first ten months of science operations, \emph{Kepler} will
survey around 2000 solar-type stars for solar-like oscillations, with
each star being observed for one month at a time.  After this initial
``Survey Phase'' approximately 100 solar-type stars will be selected
for long-term observations. At the time of writing, the number of
known solar-type oscillators has increased by approximately one order
of magnitude, thanks to \emph{Kepler}. This is with only about 40\,\%
of the total Survey Phase data available. The large homogenous sample of data presented by Kepler opens the possibility to conduct a seismic survey of the solar-type part of the colour-magnitude diagram, to compare trends in observed 
properties with trends predicted from stellar structure and evolutionary models.

In the \emph{Kepler Asteroseismic Science Consortium} (KASC) Working
Group\,\#1 has responsibility for asteroseismic analysis of solar-type
stars. First results were presented by Chaplin et al. (2010) on three
G-type stars, and many publications from the Survey Phase are planned
for the second half of 2010.

\begin{figure}
\centering
\includegraphics[width=8cm]{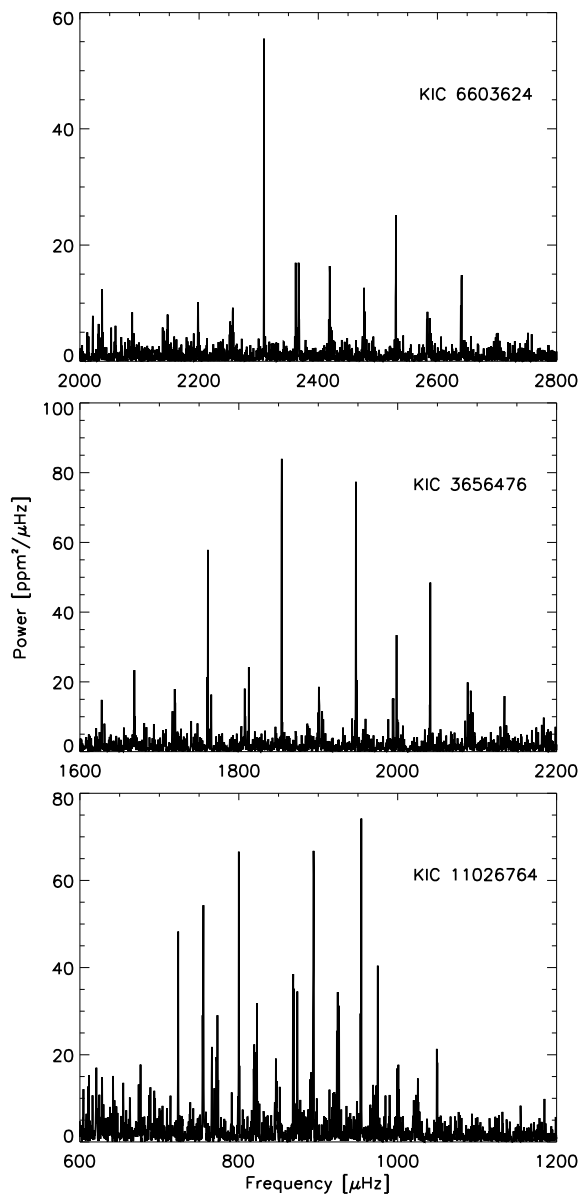}
\caption{Power density spectra of the three G-type stars analyzed by Chaplin et al. (2010)} 
\end{figure}

\section{Kepler Asteroseismic Science Consortium Working Group\,\#1: Solar-Like Oscillators}

The KASC Working Group\,\#1 is responsible for the data analysis and
modeling of the solar-type stars observed by \emph{Kepler}. The Group,
which is chaired by W.~J.~Chaplin, is divided into nine sub-groups:
\begin{itemize}

 \item [1 --] Extraction of Mean Parameters\\ chair: R. A. Garc\'ia

\item [2 --] Extraction of individual mode parameters\\ chair: T.~Appourchaux
 
\item [3 --] Analysis of Mode Excitation and Damping\\ chair: G. Houdek

\item [4 --] The Stellar Background\\ chair: C. Karoff

\item [5 --] Model Grid Comparison\\ chair: T. S. Metcalfe

\item [6 --] Fitting Models to Observed Frequencies\\ chair: M. J. P. F. G. Monteiro

\item [7 --] Modeling Rotation, Mixing and New Physics\\ chair: M. J. Thompson

\item [8 --] Analysis of Long-Term Variations\\ chair: Y. Elsworth

\item [9 --] Ground-based Follow-Up\\ chair: J. Molenda-{\.Z}akowicz

\end{itemize}

This paper gives a brief summary of the work undertaken by sub-groups
1 to 4 on the three G-type dwarfs in Chaplin et al. (2010). Metcalfe
et al. (this volume) and Molenda-{\.Z}akowicz et al. (this volume) describe,
respectively, the corresponding work performed by sub-groups 5 to 7,
and sub-group 9.

Recent improvements in the quality of asteroseismic observations, in
particular from the excellent quality CoRoT data (Michel et al. 2008),
but also from other space- and ground-based observing facilities, have
driven improvements in asteroseismic data analysis techniques. These
improvements have been followed by significant work on preparing for
the mode-parameter analysis of the Kepler data. This
analysis involves the estimation of individual and average mode
parameters, and also estimation of parameters that describe
non-resonant signatures of convection and activity that are present in
the \emph{Kepler} data. Examples include work conducted in the
framework of asteroFLAG (Chaplin et al. 2008); and work undertaken by
the CoRoT Data Analysis Team (e.g., Appourchaux et al. 2008). This has
led to the development of suites of analysis tools for application to
the \emph{Kepler} data (e.g., see Campante et al. 2010; Hekker et
al. 2010; Huber et al. 2009; Karoff et al. 2010; Mathur et al. 2010;
Mosser \& Appourchaux 2009, Roxburgh 2009). The levels of preparedness meant that
analysis of the first observations of solar-type stars by
\emph{Kepler} (see Fig.~1) could be made in a timely fashion, in order to meet the
publication deadlines set down by NASA. Shown below is a list of the
different tasks that were conducted for the Chaplin et al. (2010)
paper:
\begin{itemize}
\item[21] Oct -- Data received
\item[23] Oct -- Global seismic analysis 
\item[26] Oct -- Paper written and sent to sub-group chairs
\item[2 ] Nov -- Paper approved by working sub-group chairs\\ and sent to working group members 
\item[16] Nov --  Paper approved by working group members\\ and submitted 
\end{itemize}

\section{Signatures of convection in the stellar background}


\begin{figure}
\centering
\includegraphics[width=8cm]{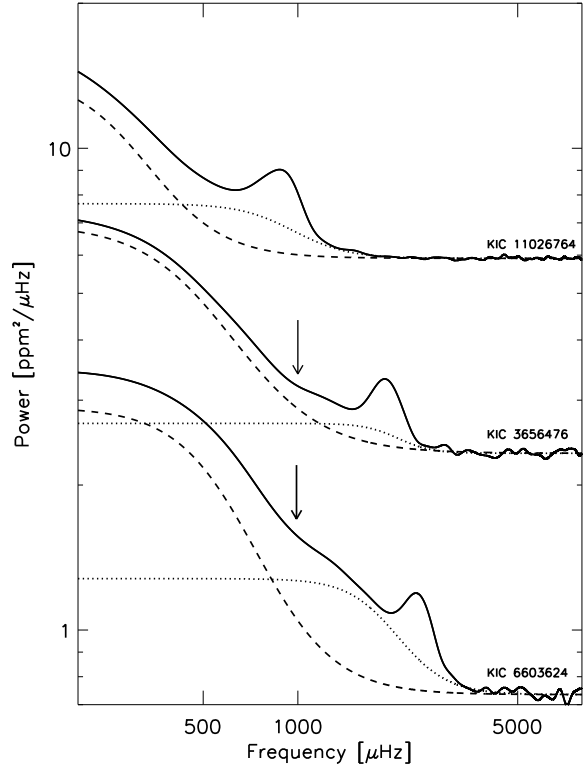}
\caption{Power density spectra of the three G-type stars analyzed by
Chaplin et al. (2010), smoothed by Gaussian running-means of width of
two times the large frequency separations. The spectra of KIC 3656476 and KIC
11026764 have been shifted upwards by 1 and 5\,ppm$^2$/$\mu$Hz, respectively. The arrows mark the locations of the signature of faculae. The dashed and dotted lines show the best-fitting models of the granulation and facular components, respectively.}
\end{figure}


Power-frequency spectra of photometric observations of the Sun and
other solar-type stars show not only signatures of oscillations, but
also signatures arising from other intrinsic stellar phenomena. In order of increasing frequency there is power due to:
rotational modulation of effects of magnetic activity, like
starspots, and also the decay of active regions; granulation; and faculae. We might also hope in the future to be able to
detect signatures of chromospheric oscillations and high-frequency
waves, both of which are observed in the Sun.

The characteristic timescales and amplitudes of the components arising
from the decay of active regions, granulation, and faculae are
commonly represented using a Harvey-like model (Harvey 1985):
\begin{equation}
B(\nu)=\sum_i{\frac{4\sigma^2_i\tau_i}{1+(2\pi\nu\tau_i)^{\alpha}}}+c,
\end{equation}
where $\sigma$ is the amplitude of the component, $\tau$ is the
characteristic timescale, $\nu$ is the frequency and $c$ is a constant that give the white noise level. The exponent $\alpha$ depends on the
``memory'' of the physical process responsible for the component.

Chaplin et al. (2010) were able to measure not only the characteristic
timescales and amplitudes of the granulation component, but also the
presence (and properties of) a component assumed to be the signature
of faculae (marked by the arrow in Fig.~2). We are now in the process of measuring the
characteristic timescales and amplitudes of the different background
components in around 200 solar-type stars observed during the first
four months of the Kepler asteroseismic survey. These stars have been
selected because they show clear the signatures of solar-like oscillations,
meaning that we will also be able to perform a full asteroseismic analysis
of their data to provide estimates of masses, radii and ages. The aim
of this study will be to identify how signatures of convection and
activity vary with stellar properties.

\section{The Echelle diagrams}

Solar-like p modes of high radial order and low angular degree are
reasonably well-described by the asymptotic relation (Tassoul 1980):
\begin{equation}
\nu_{n,l} \sim \Delta \nu(n+l/2+\epsilon)-l(l+1)D_0.
\end{equation}
Here, $n$ (the radial order) and $l$ (the angular degree) are
integers. $D_0$ is the small frequency separation parameter and $\epsilon$ is a phase constant determined by the reflection properties near the surface.

Departures of stellar oscillation frequencies from the asymptotic
relation may be shown visually by plotting the oscillation power in a so-called 
\'echelle diagram (Grec et al. 1983), as is done in Fig. 3.  Here, the oscillation power for 
each star has been plotted against the frequencies modulo the average 
large frequency separation. Individual strips of the power spectrum are 
offset vertically, such that the mean value of each \'echelle order gives the lower frequency of each \'echelle order.

Were a star to obey strictly the asymptotic relation, its frequencies
would lie in vertical ridges in the \'echelle diagram. The \'echelle
diagrams in Fig.~3 show that stars KIC~6603624 and KIC~3656476 exhibit
only small departures from an asymptotic description, whereas
KIC~11026764 shows clear deviations in its $l$~=~1 ridge. These
deviations are due to the fact that this star has started to
evolve off the main sequence and thus shows avoided
crossings (Osaki 1975; Aizenman et al. 1977). Avoided crossings result
from interactions between acoustic modes and buoyancy modes, which
affect (or ``bump'') the frequencies and also change the intrinsic
properties of the modes, with some taking on mixed acoustic and
buoyancy characteristics. The precise signatures of these avoided
crossings are very sensitive to the evolutionary state of the star. It
is therefore reasonable to assume that the presence of mixed modes
will improve significantly the age determination of stars,

For solar-type stars $\Delta \nu$ provides a measure of the inverse of
the sound travel time across the star, while $D_0$ is sensitive to the
sound-speed gradient near the core. It is conventional to define two
small frequency separations: $\delta\nu_{02}$, which is the spacing
between adjacent modes of $l$~=~0 and $l$~=~2; and $\delta\nu_{13}$,
the spacing between adjacent modes of $l$~=~1 and $l$~=~3. The
asymptotic relation then predicts that $\delta \nu_{02}$~=~$6D_0$ and
$\delta \nu_{13}$~=~$10D_0$. The spacings $\delta\nu_{02}$ are seen
clearly in all three stars.  It is normally assumed that $l$~=~3 modes
are too weak to be visible in stellar photometric observations like
the ones we have from \emph{Kepler} (Kjeldsen et al. 2008). None of
the three stars reported here shows convincing evidence for $l$~=~3
modes; however, preliminary analyses of \emph{Kepler} Survey data do
show possible evidence of $l$~=~3 modes in some stars.

We add in passing that KIC~3656476 does show signs of extra power on
the high-frequency side of its $l$~=~1 mode at $\approx 1770$ $\mu$Hz
(marked by the arrow in Fig.~3). We do not expect this power to be
due to the presence of an $l$~=~3 mode. Such power would lie on the
low-frequency side of the stronger $l$~=~1 mode, like its $l$~=~2
counterparts, which for this star clearly lie on the low-frequency
side of their $l$~=~0 neighbours. Aside from the possibility this might
be an artifact, it is conceivable that the extra power might be the
signature of a mixed mode (see also Bedding et al. 2010 for a discussion of this).


\begin{figure}
\centering
\includegraphics[width=8cm]{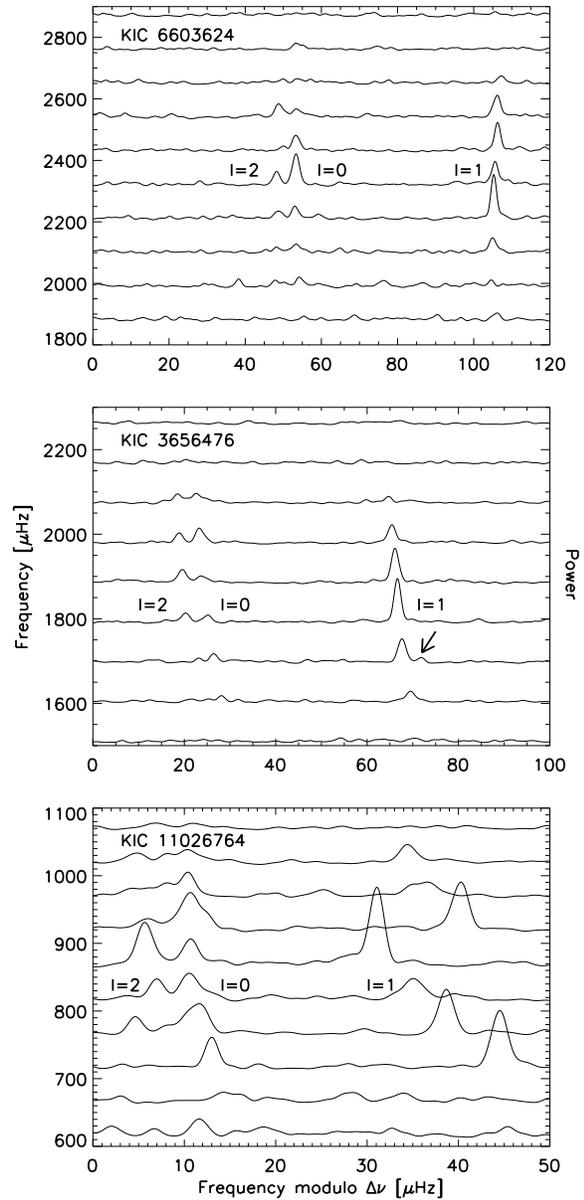}
\caption{Echelle diagrams of the three G-type stars analyzed by Chaplin
et al. (2010). The spectra have been smoothed by a Gaussian
running-mean with a width of 2 $\mu$Hz, before substrings of the
spectra were  stacked on top of one another. The large separation in the three stars were measured to 110.2 $\pm$ 0.6, 94.1 $\pm$ 0.6 and 50.8 $\pm$ 0.3 $\mu$Hz (from top to bottom).}
\end{figure}


\section{Individual mode parameters}

At the time of writing we have access to data on a few hundred
solar-type stars.  The quality of these data is such that it is
possible to extract estimates of individual frequencies, amplitudes,
and also some mode lifetimes, in a large fraction of the targets
showing evidence for solar-type oscillations. It may also be possible
to extract estimates of rotational splittings in some of the more
rapidly rotating stars.

The analysis of the three G-type stars has shown that not only can the
oscillation mode frequencies and amplitudes be measured with high
precision, but it is also possible to place constraints on the mode
lifetimes, which in all three cases appear to be similar in length to
the Sun.  Moreover, the analysis of KIC~11026764 show that the mode lifetimes plateau is at frequencies close to the frequency of maximum power $\nu_{\rm max}$ (see Fig.~4), just as for the Sun (see Chaplin et al. 2009 for a discussion of the predictions of mode lifetimes).

The observed maximum mode amplitudes of the three stars are all higher
than solar. This is in line with predictions from simple scaling
relations (Kjeldsen \& Bedding 1995; Samadi et al. 2007), which use
the inferred fundamental stellar properties as input. Data from a
larger selection of survey stars are required before we can say
anything more definitive about the relations.

Kepler will deliver multi-year datasets for the best solar-type
asteroseismic targets, and from these data we expect to be able to
extract: signatures of rapid structural changes in the stellar
interiors, from the borders of convective regions and from zones of
ionization in the near-surface layers; rotational splittings as a
function of $n$ and $l$, and possibly subtle signatures originating
from differential rotation; and changing oscillation mode frequencies
and amplitudes due to stellar cycles (see Karoff et al. 2009 for
details)


\begin{figure}
\centering
\includegraphics[width=8cm]{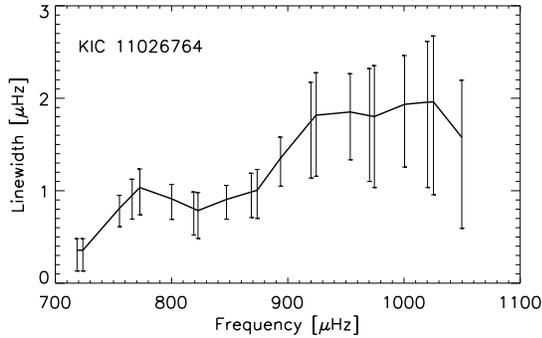}
\caption{Oscillation mode linewidth for KIC~11026764. Note how similar the change in linewidth as a function of frequency is to what has been observed for the Sun.}
\end{figure}


\section*{Acknowledgments}

CK acknowledges financial support from the Danish Natural Sciences
Research Council.

\end{document}